\documentclass[12pt]{revtex4-1}
\usepackage{fullpage}
\usepackage{graphicx,graphics}
\usepackage[sort&compress]{natbib}

\usepackage{setspace}
\usepackage{amssymb}
\usepackage[usenames,dvipsnames]{color}

\begin{document}

\title{Modeling elastic properties of polystyrene through coarse-grained molecular dynamics simulations}
\author{Yaroslav Beltukov}
\affiliation{Ioffe Institute, Politechnicheskaya Str. 26, 194021 St. Petersburg, Russia}
\author{Igor Gula}  
\affiliation{Department of Physics, Chemistry, and Pharmacy, University of Southern Denmark, DK-5230 Odense M, Denmark}
\author{Alexander M. Samsonov}
\affiliation{Ioffe Institute, Politechnicheskaya Str. 26, 194021 St. Petersburg, Russia}
\author{Ilia A. Solov'yov}
\email{ilia@sdu.dk}
\affiliation{Ioffe Institute, Politechnicheskaya Str. 26, 194021 St. Petersburg, Russia}
\affiliation{Department of Physics, Chemistry, and Pharmacy, University of Southern Denmark, DK-5230 Odense M, Denmark}

\begin{abstract}
This paper presents an extended coarse-grained investigation of the elastic properties of polystyrene. In particular, we employ the well-known MARTINI force field and its modifications to perform extended molecular dynamics simulations at the $\mu$s timescale, which take slow relaxation processes of polystyrene into account, such that the simulations permit analyzing the bulk modulus, the shear modulus, and the Poisson ratio. We show that through the iterative modification of MARTINI force field parameters it turns out to be possible to affect the shear modulus and the bulk modulus of the system, making them closer to those values reported in the experiment.
\end{abstract}

\maketitle

\newpage
\section{Introduction}

Polymers and polymer-based materials are widely used as components of various products and goods nowadays and are employed in daily life. Such an extensive use is mainly due to the wide range of physical and mechanical properties of polymers, which can be enhanced by numerous additives~\cite{camargo2009nanocomposites,Mark1999,Mark2013,Harper2002}. Modern technologies have made it possible to create new types of polymer composites using nanoparticles of various nature and composition \cite{fawaz2015synthesis}. The manufacturing process is, however, often associated with specific technological difficulties, including the monitoring of the aggregation-disaggregation processes to obtain the uniform distribution of nanoparticles in the polymer matrix~\cite{supova2011effect}.

Preparation, characterization, and applications of polymeric nanocomposites with various nanoparticles has influenced multiple research areas~\cite{nasir2015review,armstrong2015introduction}. Nanocomposites with silica nanoparticles have particularly attracted substantial academic and industrial interest~\cite{zou2008polymer,mathioudakis2016multiscale,rahman2012synthesis}; polymer/silica composites are seemingly the most commonly reported ones, which have also been successfully employed in a variety of applications, including tire industry, food industry, civil engineering, and recyclable materials \cite{Mark2013,Harper2002}.

Different polymers have been used for polymer/silica composites. In particular polypropylene~\cite{bracho2012functionalization}, polyimide~\cite{musto2004polyimide}, polyamide~\cite{sarwar2008polyamide}, poly(ethylene terephthalate)~\cite{liu2004preparation}, polymethylmethacrylate~\cite{yang2004pmma}, polyurethane~\cite{petrovic2000structure} have been employed, while polystyrene was also a popular focus of both experimental~\cite{liu2005thermal, vaziri2011thermophysical, bartholome2005viscoelastic} and theoretical~\cite{mathioudakis2016multiscale, ndoro2011interface, ghanbari2011interphase} research. Polystyrene is a general-purpose, well-known frequently utilized, light weight, famous plastic with worthy dimensional stability, noble chemical resistance, easy processing, and low cost \cite{Mark1999}. It has been extensively utilized in packing materials, electronics, household appliances, etc.

Nowadays, computer simulation becomes a promising and powerful tool to investigate the structure and properties of polymers, thereby providing an in-depth understanding of experimental data.
Historically, polymers have been widely studied through theoretical continuous models \cite{Rivlin1948,Mooney1940,LL7} as well as discreet microscopic models \cite{WallFlory1951,JamesGuth1943ThrElstPropRub}, that permit establishing bulk properties of the materials. The well-known hyperelasticity model of Mooney and Rivlin (MR) \cite{Rivlin1948,Mooney1940}, and first microscopical models for polymer networks elasticity were suggested in the 50th. Since then generalization of the MR model was suggested, but the first successful microscopical / molecular models, that take entanglements effects into account, are available since 80th \cite{Edwards1967StatMechTopologyI,Gaylord1982,MergellEveraers2001TubeModels,RubinPanyuk2002ElastPolymNetw}.
Molecular dynamics (MD) was also seen as an important tool to study polymer properties.
The most popular approach exploited in MD simulations of polymer dynamics is the generic bead-spring model of Kremer and Grest (KG) \cite{GrestKremer1986MDSimPolymHeatBath}, originally invented for polymer melts. The KG model is actively applied until nowadays to study the dynamics of melts  \cite{KremerGrest1990DynEntanglLinPolyMelt,SvaneborgKarimiHojdisFleckEveraers2016MultiscaleApproach}, solid rubbery polymers \cite{GrestKremer1990StatPropRandXLnkdRubb,EveraersKremer1996ElastPropPolyNetw,SvaneborgEveraersGrestCurro2008StressContributions} and glassy polymers after proper modification \cite{Bennemann1998}.

Recently, due to the emerging supercomputer facilities, it also became feasible to study polymers through advanced MD simulations, including atomistic MD and unprecedentedly long simulation on experimentally accessible timescales. Atomistic MD has been widely applied to study biological polymers, such as, e.g.,  proteins~\cite{Dubey2018,Kimo2018,Kattnig2018a,Akimov2018,Friis2017} and DNA~\cite{Klecka2018,Jepsen2017a,Zou2012a}, and additionally it was also applied to synthetic polymers, such as, e.g. polystyrene \cite{rossi2011coarse,Chakravarthy2000}. Advanced MD simulation protocols have been developed for both explicit atom (EA) models, treating every atom as a separate interaction site \cite{Yoon1993}, but also united atom (UA) models where several atoms are grouped into a single bead  (e.g., CH$_2$ or CH$_3$)~\cite{barrat2010molecular}. The reduction of the number of interaction sites in the UA model leads to significant simplifications of the complex polymeric systems, and, therefore, permits performing significantly longer simulations capable of predicting many universal polymer properties independent of local chemical structure.

UA model has been successfully employed for studying polystyrene~\cite{lyulin2002correlated,lyulin2002molecular,lyulin2003molecular,lyulin2005strain,lyulin2007time}. In these coarse-grained simulations the typical simulation time was about 2~ns, and it was observed that polystyrene chains experience several relaxation processes with very different relaxation times~\cite{lyulin2007time}. Some of the established relaxation times turned out to be comparable to the overall simulation time, which indicates the necessity of significantly longer simulations. There are alternatives to the UA approach for studying polystyrene on a coarse-grained level, which includes the so-called 2:1 model~\cite{harmandaris2009dynamics} and the MARTINI force field~\cite{Marrink2007,Marrink2013,rossi2011coarse}.

The paper presents extended coarse-grained MD simulations of polystyrene and aims to establish its various elastic properties, such as the bulk modulus, the shear modulus, and the Poisson ratio computationally. The simulations rely on the well-established MARTINI force field \cite{Marrink2007,Marrink2013,rossi2011coarse}, which is revealed to be somewhat inaccurate in predicting all the first order elastic moduli of polystyrene simultaneously. Investigations reported here demonstrate that through iterative modification of the force field parameters it turns out to be possible to affect the shear modulus and the bulk modulus, making them closer to those values reported in the experiment. Through these modifications, it is revealed that the MARTINI approximation (and its modifications) can provide a qualitative force field for computing of the first-order elastic constants.
The key aim of this study is to provide a model and a force field that have been justified to be qualitatively applicable for studying pure polystyrene, such that this model could be further extended through doping with nanocomposites such as, e.g. the SiO$_2$ nanoparticles.

The paper delivers a general recipe on how to use MD to obtain elastic properties of polymers. It discusses several theoretical approaches that can be used to determine the elastic moduli and argues about their accuracy and applicability. Specifically, the analysis of potential energy and internal pressure of the polystyrene under strain has been analyzed, and it has been demonstrated that both approaches may yield significantly different results in terms of the elastic moduli values and polystyrene relaxation time.

\section{Theoretical methods}

This section discusses the key theoretical and computational methods used for studying the elastic properties of polystyrene.
Below we describe the general approach of the computational method used in the investigation, explain how the simulations were set up in the coarse-grained representation and how the interactions between coarse-grained beads were modeled. Finally, a table summarizing all the performed simulations is presented.

\subsection{System preparation: all atom representation}

\begin{figure}[!t]
\centering
\includegraphics[width=15cm]{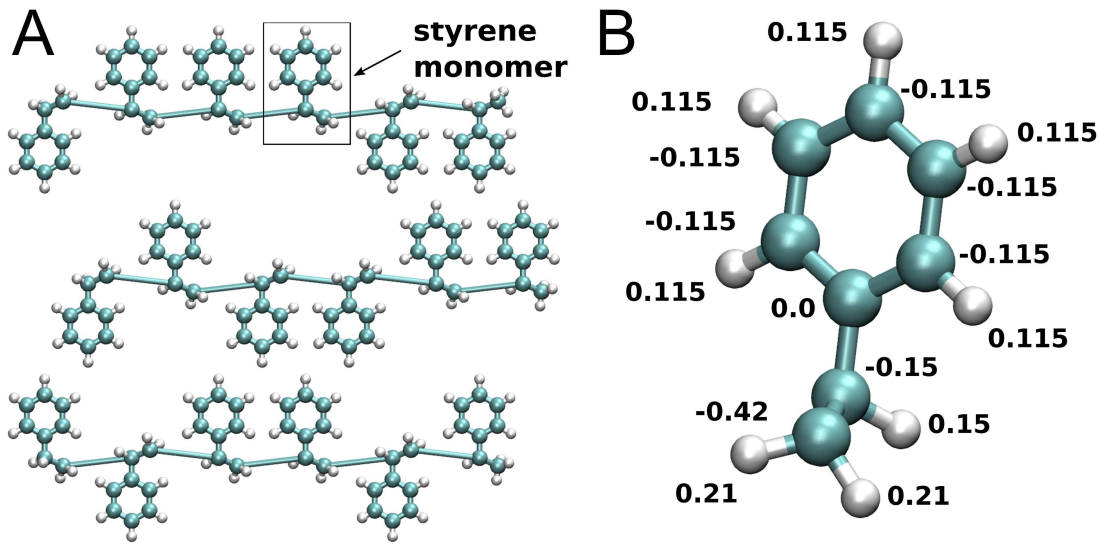}
\caption{
\textbf{Atomistic model of polystyrene.} \textbf{A:} Illustration of atactic chains used to construct the bulk polystyrene structure. Styrene monomers were placed randomly from two sides of the polymer chain. \textbf{B:} Styrene monomer consists of 8 carbon atoms (colored in cyan) and 8 hydrogen atoms (colored in white). The partial charges of the atoms were adopted from an earlier study \cite{Vanommeslaeghe2010} and were used in the atomistic simulations.
\label{fig:PS_atomistic}
}
\end{figure}

The bulk polystyrene studied here was initially constructed in the all-atom representation. The system was assumed to contain 216 polystyrene strings consisting of 120 monomers each. The strings were placed in a simulation box of 167.2~\AA$\times$167.2~\AA$\times$167.2~\AA\ to ensure the volumetric density of the system being equal to 0.96 g/cm$^3$, being consistent with the literature \cite{Hocker1971,Zoller1995,rossi2011coarse}. Initially, styrene monomers were placed randomly from the two sides of the parent polymer chain, as illustrated in Fig.~\ref{fig:PS_atomistic}A. The studied model assumed several short polystyrene strings instead of a single long chain to ensure faster relaxation of the system. The studied model is consistent with earlier investigations \cite{rossi2011coarse,Mathioudakis2016}, where bulk properties of polystyrene were probed on a system consisting of multiple shorter polymeric chains.

Atomistic MD simulations of the constructed bulk polymer were performed to ensure initial equilibration of the system prior to coarse-graining. MD program NAMD 2.12 \cite{PHIL2005} was used in the computations, while interactions between the atoms of the styrene monomers were modeled with the CGenFF force field \cite{Vanommeslaeghe2010}; the partial charges of one monomer are indicated in Fig.~\ref{fig:PS_atomistic}B. Analysis of the simulations results and visualization of molecular structures were accomplished with VMD 1.9.2 \cite{HUMP96}. Simulations were carried out with the 1~fs integration time step. The cut-off distances for the van der Waals and Coulomb interactions were set to 12~\AA\,, where the long-range electrostatic interactions were treated using the PME method \cite{darden1993particle,Solovyov2012c}, employing periodic boundary conditions. The NVT statistical ensemble was used for the simulation, with a temperature of 300~K by applying Langevin forces with a damping coefficient of 5~ps$^{-1}$. The system was simulated in all-atom representation for 100~ns to ensure a random and uniform polystyrene configuration prior coarse-graining, while the elastic properties were computed within the coarse-grained approximation. This study aims to fine-tune the parameters of the coarse-grained potential in such a way that it can reproduce experimental observables directly, i.e., without transferring the coarse-grained model to an all-atom one, as done in some other studies \cite{ZhangMoreiraStuehnDaoulasKremer2014EquilHierarchStrat,SvaneborgKarimiHojdisFleckEveraers2016MultiscaleApproach}.

\subsection{Coarse-graining the system}

\begin{figure}[!h]
\centering
\includegraphics[width=15cm]{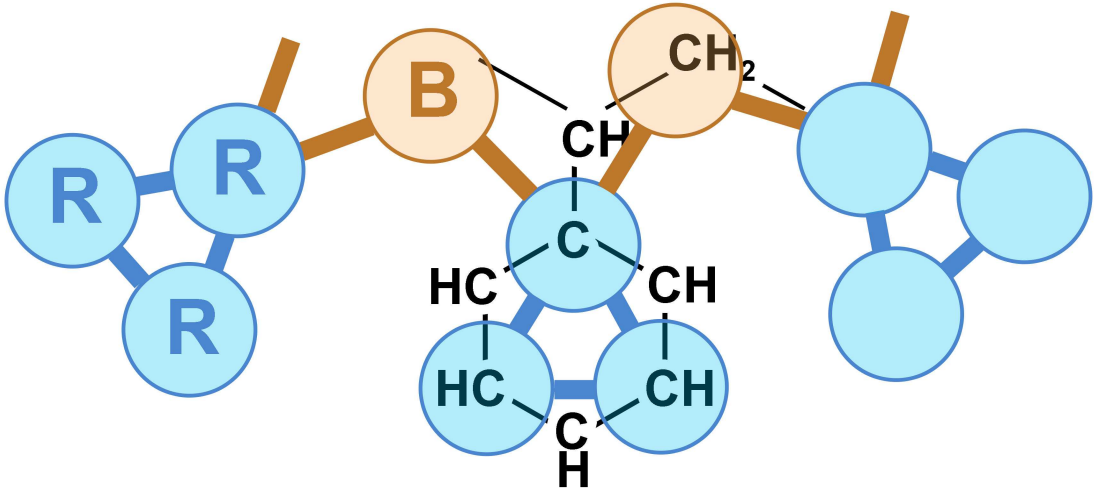}
\caption{
\textbf{Coarse-grained model of polystyrene.} The atomistic and coarse-grained models of polystyrene overlayed atop each other. The labels of the coarse-grained beads were used in the simulations and their corresponding parameters are summarized in Table~\ref{tab:parameters}.
\label{fig:coarse_grained_def}
}
\end{figure}

The polystyrene chains were coarse-grained to permit longer simulations and ensure equilibration of the system. Each styrene monomer was substituted with four coarse-grained beads, as illustrated in Fig.~\ref{fig:coarse_grained_def}. Such a representation is consistent with earlier studies \cite{rossi2011coarse}; the backbone carbon atoms of the polystyrene are modeled through one bead of type B, while three beads of type R represent the styrene side chain.

The coarse-grained system of bulk polystyrene was constructed from the last frame of the atomistic MD simulation, such that each of the atomistic polystyrene chains was represented by a coarse-grained analog, using an in-house conversion script.

\subsection{Coarse-grained molecular dynamics and MARTINI force field}

\begin{table}[!t]
\centering
\begin{tabular}{|c|c|c|}
\hline
Bonded     & $r_0$ (\AA) & $k_0$ (kcal/mol/\AA$^2$)\\
\hline
B--R & 2.7         & 9.5602\\
R--R & 2.7         & constrained\\
\hline
\hline
Angular     & $\vartheta_0$ (degrees) & $k_{\vartheta}$ (kcal/mol)\\
\hline
B--R--B & 52                  & 65.7266\\
B--B--R & 120                 &  2.9876\\
B--R--R & 120                 & 11.9503\\
R--R--R & 60                  & 0.0\\
\hline
\hline
Nonbonded pairwise     & $r_{min}$ (\AA) & $\varepsilon$ (kcal/mol)\\
MARTINI-std & & \\
\hline
B $\cdots$ B & 4.82659                 & -0.6274\\
R $\cdots$ R & 4.60209                 & -0.5736\\
R $\cdots$ B & 4.82659                 & -0.5557\\
\hline
%Nonbonded pair     & $r_{min}$ (\AA) & $\varepsilon$ (kcal/mol)\\
MARTINI-1.5 & & \\
\hline
B $\cdots$ B & 4.82659                 & -0.4183\\
R $\cdots$ R & 4.60209                 & -0.3824\\
R $\cdots$ B & 4.82659                 & -0.3705\\
\hline
%Nonbonded pair     & $r_{min}$ (\AA) & $\varepsilon$ (kcal/mol)\\
MARTINI-2 & & \\
\hline
B $\cdots$ B & 4.82659                 & -0.3137\\
R $\cdots$ R & 4.60209                 & -0.2868\\
R $\cdots$ B & 4.82659                 & -0.2778\\
\hline
%%Nonbonded pair     & $r_{min}$ (\AA) & $\varepsilon$ (kcal/mol)\\
%MARTINI-3 & & \\
%\hline
%B $\cdots$ B & 4.82659                 & -0.2091\\
%R $\cdots$ R & 4.60209                 & -0.1912\\
%R $\cdots$ B & 4.82659                 & -0.1852\\
%\hline
\end{tabular}
\caption{Parameters of bonded, angular and nonbonded interactions employed in the coarse-grained simulations. Notations of the beads are consistent with \cite{rossi2011coarse} such that B and R indicate the backbone and ring, respectively. The parameters for the standard MARTINI force field for polystyrene have been adopted from an earlier study \cite{rossi2011coarse} and are denoted as MARTINI-std in the table and text. Two modifications to the force field are also considered where the non-bonded energy parameter $\varepsilon$ of the standard force field is scaled by 1.5 and 2; the corresponding force fields are denoted as MARTINI-1.5 and MARTINI-2, respectively.
\label{tab:parameters}}
\end{table}

The coarse-grained bulk polystyrene was simulated dynamically employing NAMD 2.12 \cite{PHIL2005} in the NVT statistical ensemble. In production simulations, i.e., those that were used for data sampling, the temperature $T$ was assumed 300~K, the number of coarse-grained particles was 103,680, while the volume of the simulation box was different, depending on the simulation; all performed simulations are listed in the summarizing Table~\ref{tab:protocol}. Prior to the production simulations, an extensive equilibration simulation of the system was carried out to ensure the uniform density distribution of the polymers in the system. The temperature was gradually changed in the equilibration simulations from 1000~K to 300~K as indicated in Table~\ref{tab:protocol} to anneal the system to thermal equilibrium.

The equation of motion in the case of the coarse-grained dynamics is the same as in the case of atomistic simulations and reads

\noindent
\begin{equation}
m_i{\vec a}_i=m_i\frac{d^2{\vec r}_i}{dt^2}={\vec F}_i - \gamma{\vec v}_i + \sqrt{2 k_{\mathrm{B}} T\gamma} {\vec R}_i(t),\ \ \ \ \forall i=1\ldots N_{beads}.
\label{eq:LangevinDynamicsEquation}
\end{equation}

\noindent
Here ${\vec F}_i$ is the force acting on a coarse-grained bead $i$, $k_{\mathrm{B}} T$ denotes the thermal energy in the system, $\gamma$ is the damping coefficient that is assumed equal 1~ps$^{-1}$ in all coarse-grained simulations and ${\vec R}_i(t)$ represents a delta-correlated stationary Gaussian process with zero mean. Equation~(\ref{eq:LangevinDynamicsEquation}) is called as the Langevin equation \cite{Solovyov2012c,PHIL2005} and is used to control the temperature in the system. The difference from the atomistic simulations is that it is applied to every bead instead of every atom in the system. Due to the significantly higher mass of one bead, Eq.~(\ref{eq:LangevinDynamicsEquation}) can be solved efficiently with a higher integration time step, as compared to the atomistic simulations; a time step of 15~fs was used in the present investigation.

In the performed simulations, the forces acting between the beads are determined by the parametric MARTINI potential \cite{Marrink2007,Marrink2013,rossi2011coarse}, which reads

\begin{eqnarray}
\nonumber
U&=&-\frac{1}{2}\sum_{i\ne j}\varepsilon\left(\left(\frac{r_{min}}{r_{ij}}\right)^{12}-2\left(\frac{r_{min}}{r_{ij}}\right)^{6}\right)+\\
\label{eq:MARTINI}
&+&\sum_{bonded}k_{0}(r_{ij}-r_0)^2+\frac{1}{2}\sum_{angular}k_{\vartheta}(\cos(\vartheta_{ijk})-\cos(\vartheta_0))^2.
\end{eqnarray}

This simple parametrization mimics the non-bonded van der Waals interactions between the beads through the Lennard-Jones potential and the covalent interactions through a number of parameters and fit functions.  The first term in Eq.~(\ref{eq:MARTINI}) is the non-bonded term. The second and the third terms describe the potential energy arising due to stretching of bonds between pairs of bound beads, and the change of angles between every topologically defined triple of beads in the system. The force field parameters $r_0$, $k_0$, $\vartheta_0$, $k_{\vartheta}$. $r_{min}$, and $\varepsilon$ for styrene monomers are available from the literature \cite{rossi2011coarse} and, therefore, have also been utilized in the present study. For convenience, Table~\ref{tab:parameters} lists these parameters for all the relevant pairs and triples of the coarse-grained beads R and B, see Fig.~\ref{fig:coarse_grained_def}. The table includes the parameters for the standard MARTINI potential (MARTINI-std), while at the same time it also summarizes two sets of non-bonded parameters (MARTINI-1.5 and MARTINI-2), which have been employed in the present investigation to fine-tune the interactions.

\subsection{Simulation protocol}

Equations~(\ref{eq:LangevinDynamicsEquation})-(\ref{eq:MARTINI}) were solved numerically, where the simulation protocol was following the standard guidelines advised by NAMD \cite{PHIL2005}. For the coarse-grained simulations, the cutoff interaction distance was set to 12~\AA, the switch distance to 9.0~\AA, the pair list distance 14.0~\AA, and the dielectric constant was assumed equal to 15.

\begin{table}[!h]
\centering
\begin{tabular}{|c|c|c|c|c|}
\hline
Simulation type     & Box size                      & Simulation  & Temperature & Force fields \\
                    & (\AA$\times$\AA$\times$\AA)   & length (ns) & (K)         & employed\\
\hline
equilibration          & 167.2$\times$167.2$\times$167.2     & 30 & 300 &  MARTINI-std\\
equilibration          & 167.2$\times$167.2$\times$167.2     & 100 & 1000 &  MARTINI-std\\
equilibration          & 167.2$\times$167.2$\times$167.2     & 100 & 800 &  MARTINI-std\\
equilibration          & 167.2$\times$167.2$\times$167.2     & 100 & 600 &  MARTINI-std\\
equilibration          & 167.2$\times$167.2$\times$167.2     & 100 & 400 &  MARTINI-std\\
equilibration          & 167.2$\times$167.2$\times$167.2     & 2800 & 300 &  MARTINI-std\\

\hline
volumetric          & 171.0$\times$171.0$\times$171.0     & 800 & 300 &   MARTINI-1.5,2\\
volumetric          & 171.5$\times$171.5$\times$171.5     & 800 & 300 &   MARTINI-std,1.5,2\\
volumetric          & 172.0$\times$172.0$\times$172.0     & 800 & 300 &   MARTINI-std,1.5,2\\
volumetric          & 172.5$\times$172.5$\times$172.5     & 800 & 300 &   MARTINI-std,1.5,2\\
volumetric          & 173.0$\times$173.0$\times$173.0     & 800 & 300 &   MARTINI-std,1.5,2\\
volumetric          & 173.5$\times$173.5$\times$173.5     & 800 & 300 &  MARTINI-std,1.5,2\\
volumetric          & 174.0$\times$174.0$\times$174.0     & 800 & 300 &   MARTINI-std,1.5,2\\
volumetric          & 174.5$\times$174.5$\times$174.5     & 800 & 300 &   MARTINI-std\\
\hline
axial               & 170.5$\times$172.0$\times$172.0     & 800 & 300 &   MARTINI-std,1.5,2\\
axial               & 171.0$\times$172.0$\times$172.0     & 800 & 300 &   MARTINI-std,1.5,2\\
axial               & 171.5$\times$172.0$\times$172.0     & 800 & 300 &   MARTINI-std,1.5,2\\
axial               & 172.0$\times$172.0$\times$172.0     & 800 & 300 &   MARTINI-std,1.5,2\\
axial               & 172.5$\times$172.0$\times$172.0     & 800 & 300 &   MARTINI-std,1.5,2\\
axial               & 173.0$\times$172.0$\times$172.0     & 800 & 300 &   MARTINI-std,1.5,2\\
axial               & 173.5$\times$172.0$\times$172.0     & 800 & 300 &   MARTINI-std,1.5,2\\
\hline
\end{tabular}
\caption{A summary of the performed coarse-grained simulations reflecting two compression regimes (volumetric and axial) studied. Extensive equilibration was also performed prior to the production simulation, and is summarized in the table.
\label{tab:protocol}}
\end{table}

A series of simulations were carried out to determine the elastic properties of the bulk polystyrene and to establish the optimal interactions within the system; the complete list of simulations is summarized in Table~\ref{tab:protocol}. Simulations were performed for different sizes of the simulation box, as indicated in the table, and with different force fields for treating the non-bonded interactions, see Table~\ref{tab:parameters}. Two compression regimes were considered: uniform volumetric compression and axial compression. These perturbations of the bulk polystyrene were modeled by changing the size of the simulation box while keeping the same number of particles inside. The resulting changes in energy and internal pressure were then recorded and analyzed, as discussed in the following sections.

\section{Results and Discussion}

This section presents and discusses the results of the simulations. Bulk polystyrene is first equilibrated in a cubic simulation box to ensure a uniform density distribution. Elastic properties are computed next for the equilibrated system for the different variants of the MARTINI coarse-grained potential, studied here. The obtained results are compared with earlier investigations obtained both experimentally and computationally.

\subsection{Equilibration of polystyrene}

\begin{figure}[!h]
\centering
\includegraphics[width=15cm]{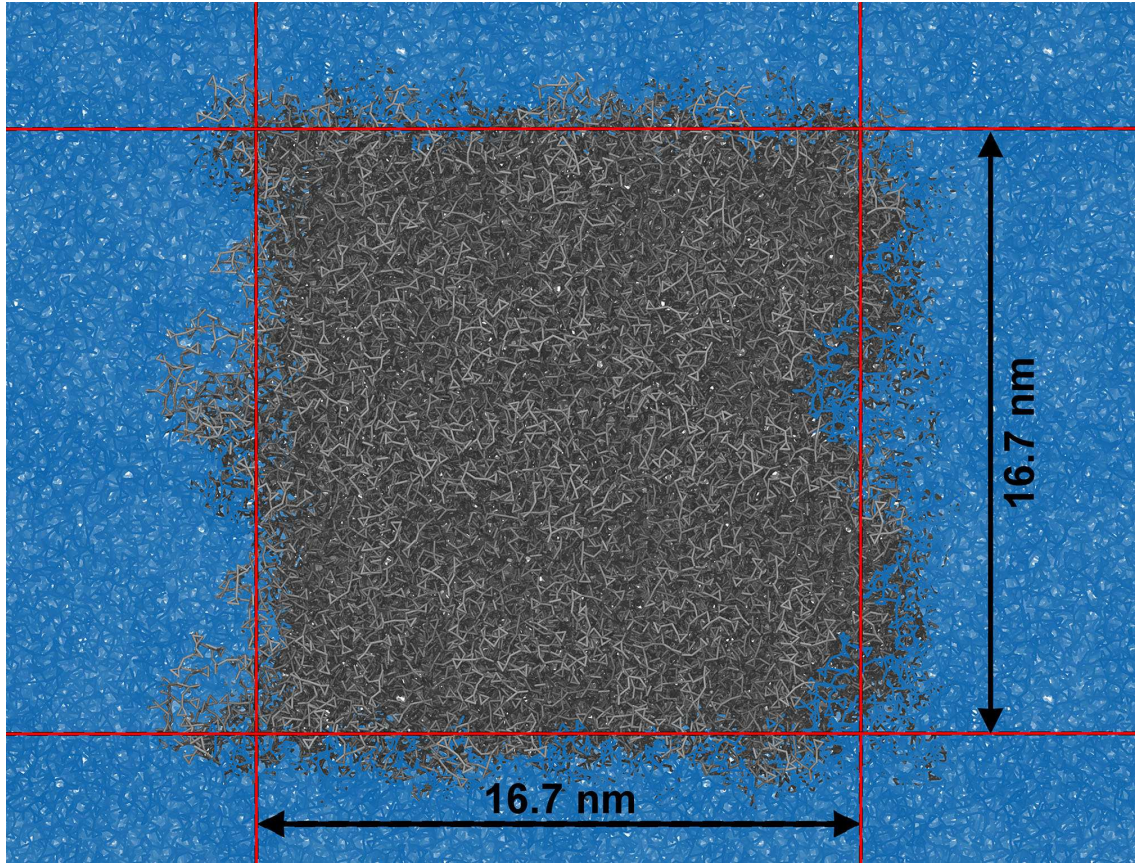}
\caption{
\textbf{Equilibrated polystyrene.} A simulation box filled with polystyrene chains as obtained after the extended equilibration, see Table~\ref{tab:protocol}. Periodic boundary conditions are indicated with red lines, while the ``real'' polymers are shown in gray.
\label{fig:postEquil}
}
\end{figure}

The density of polystyrene varies depending on its type and manufacturing technique. A characteristic value is expected to be around 960 kg/m$^3$ \cite{Hocker1971,Zoller1995,rossi2011coarse}, which is also assumed in the present investigation. The edge length of the simulation box can be determined from the expected density value $\rho$ as:

\begin{equation}
L=\left(\frac{M_{sty}N_{0}N_{str}}{\rho}\right)^{1/3},
\label{eq:edgeLength}
\end{equation}

\noindent
where $M_{sty}=8(m_{H}+m_{C})$ is the mass of one styrene monomer $N_{0}=120$ is the number of monomers in one polystyrene string, while $N_{str}=216$ is the number of strings in the simulation box. For the desired value of the polystyrene density, the size of the cubic simulation box becomes 167.2$\times$167.2$\times$167.2 \AA$^{3}$, see Table.~\ref{tab:protocol}. Extensive equilibration has been performed following a protocol that includes temperature annealing, as outlined in the table. The equilibration simulation was 3.23 $\mu$s long and allowed to ensure a uniform density distribution of bulk polystyrene. Figure~\ref{fig:postEquil} shows a snapshot of the system after the equilibration and illustrates how periodic boundary conditions were effectively used to eliminate the possible boundary artifacts. In other words, the simulated system represents an infinite sample of polystyrene that has been severely intertwined following the annealing procedure. Indeed, Fig.~\ref{fig:postEquil} features an utterly random orientation of the polystyrene monomers without any preferred orientation.

\begin{figure}[!t]
\centering
\includegraphics[width=13cm]{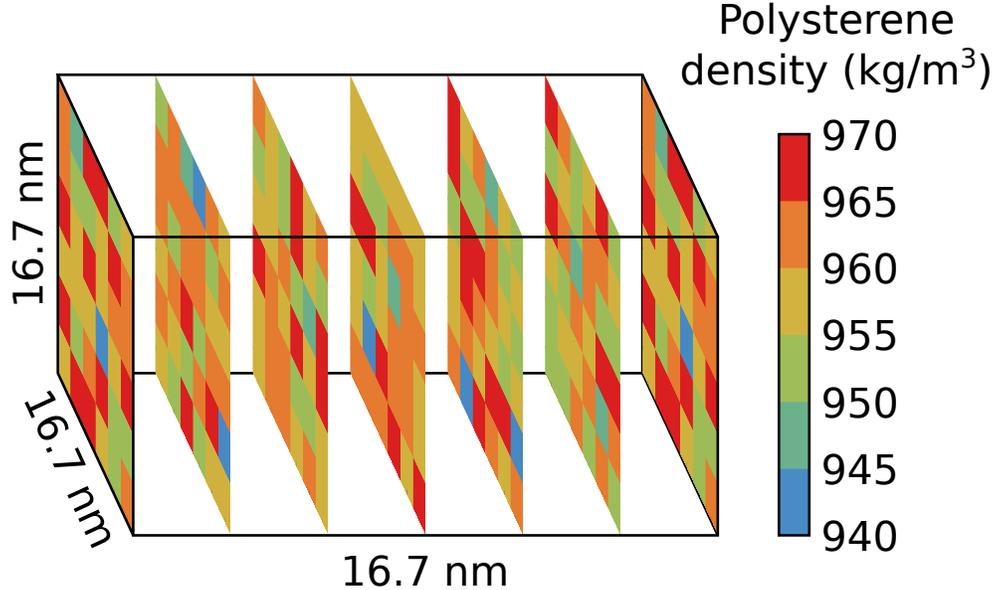}
\caption{
\textbf{Density of the equilibrated polystyrene.} The simulation box with polystyrene, as shown in Fig.~\ref{fig:postEquil} was subdivided into 6$\times$6$\times$6 cubic cells, where the average density of the contained material was computed. The density for each cube was defined according to Eq.~(\ref{eq:density}) and the averaging was performed over 20 frames taken from the equilibration MD simulation.
\label{fig:density}
}
\end{figure}

Figure~\ref{fig:density} shows the average density of the bulk polystyrene inside the simulation box. Here the simulation box was subdivided into 6$\times$6$\times$6 cells, and the average density of the system was computed in each of these cells following the definition:

\begin{equation}
\langle\rho_{cell}\rangle=\frac{1}{N_{conf}}\sum_{t=1}^{Nconf}\sum_{i\in cell}m_i/V_{cell},
\label{eq:density}
\end{equation}

\noindent
where $m_i$ is the mass of an atom inside a cell (which is calculated from the corresponding coarse-grained bead), while $V_{cell}$ is the volume of the cell. $N_{conf}$ is the number of snapshots of the simulation box used for averaging. In the present calculations we have assumed $N_{conf}=20$.

Figure~\ref{fig:density} features a three-dimensional density distribution that does not show any specific regions with significantly different density. The distribution is rather uniform and illustrates that small local fluctuations of density are possible due to the small and finite size of the sampling cells. Figure~\ref{fig:density} demonstrates that the extended equilibration was sufficient to achieve a uniform distribution of the system, making it appropriate for the study of elastic properties.

%%%%%%%%%%%%%%%%%%%%%%%%%%%%%%%%%%%%%%%%%%%%%%%%%%%%%%%
\subsection{Volumetric compression}
%%%%%%%%%%%%%%%%%%%%%%%%%%%%%%%%%%%%%%%%%%%%%%%%%%%%%%%%%%%%%%%%%%

The first compression regime to consider is the volumetric compression of the bulk polystyrene. In this case, the equilibrated system is subjected to a uniform compression applied to all sides of the simulation box. This is achieved by varying the size of the simulation box, which effectively results in the change of the polystyrene density, and internal stress, while the performed change in the box size is easily related to strain. As the applied strain is finite, and the compression of the system is hydrostatic, it is possible to employ the Birch-Murnaghan isothermal equation of state \cite{Birch1947,Murnaghan1944} to describe the relationship between the volume of the simulation box and the internal pressure to which it is subject. Specifically, the third-order Birch–Murnaghan isothermal equation of state can be used and reads as:

\begin{equation}
P(V)=\frac{3K}{2}\left[\left(\frac{V_0}{V}\right)^{7/3}-\left(\frac{V_0}{V}\right)^{5/3}\right]\left\{1+\frac{3}{4}(K^{\prime}-4)\left[\left(\frac{V_0}{V}\right)^{2/3}-1\right]\right\}
\label{eq:BirchMurnagham}
\end{equation}

\noindent
where $P$ is the pressure, $V_0$ is the reference volume of the simulation box, $V$ is the deformed volume, $K$ is the bulk modulus, and $K^{\prime}$ is the derivative of the bulk modulus with respect to pressure. Equation~(\ref{eq:BirchMurnagham}) allows to compute the internal (potential) energy of bulk polystyrene as the function of volume, by integration:

\begin{equation}
E(V)=E_0+\frac{9V_0K}{16}\left\{\left[\left(\frac{V_0}{V}\right)^{2/3}-1\right]^3K^{\prime}+\left[\left(\frac{V_0}{V}\right)^{2/3}-1\right]^2\left[6-4\left(\frac{V_0}{V}\right)^{2/3}\right]\right\}.
\label{eq:BirchMurnaghamEnergy}
\end{equation}

\noindent
Here $E_0$ is the integration constant which has the meaning of the reference energy of the polystyrene sample. This equation of state can be used to determine the bulk modulus $K$ if the change of the internal energy of the system is known as the function of its volume, which in turn can be established from simulations.

\begin{figure}[!t]
\centering
\includegraphics[width=15cm]{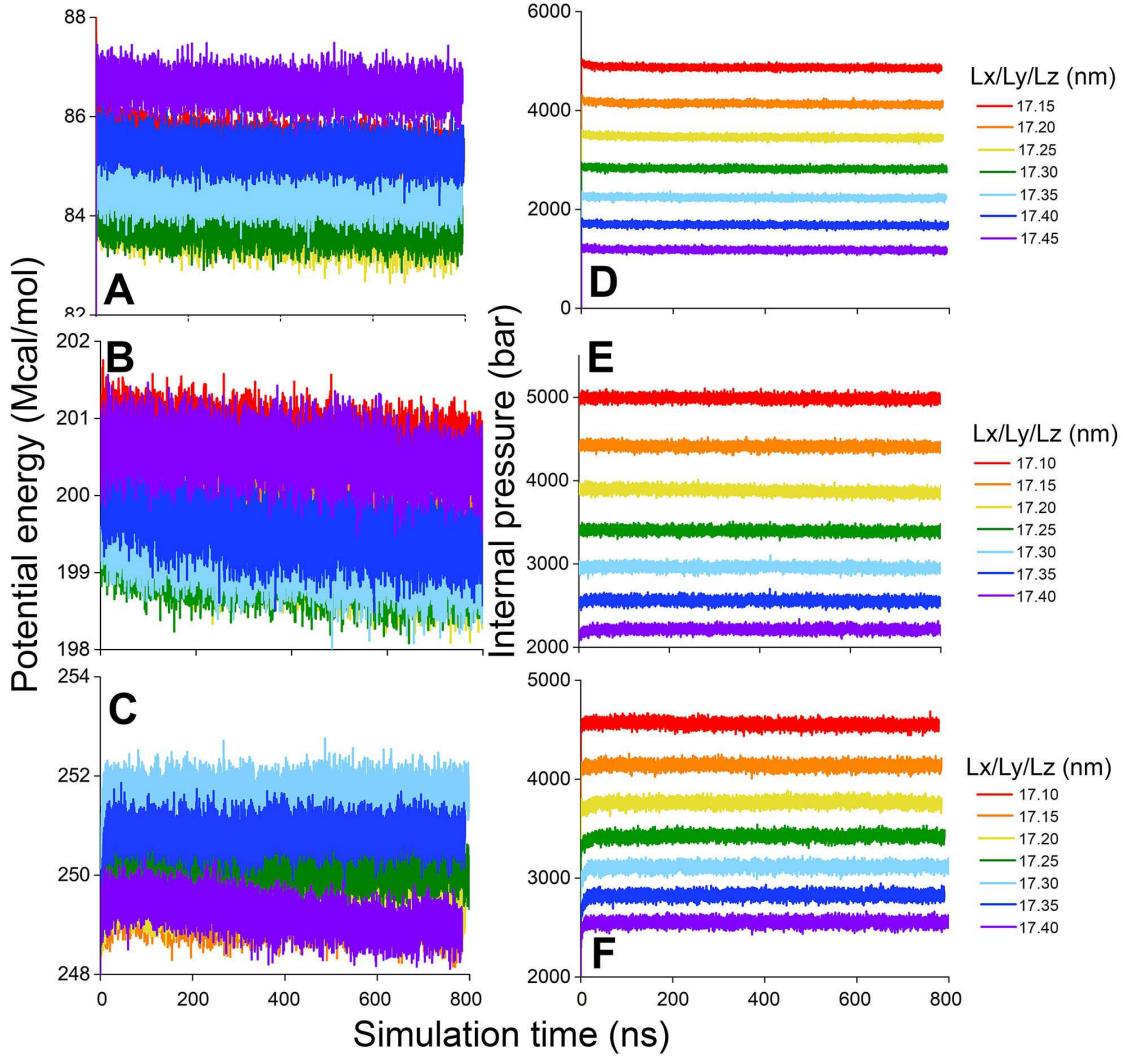}
\caption{
\textbf{Characterizing volumetric compression of the polystyrene sample.} Time evolution of the potential energy (\textbf{A-C}) and internal pressure (\textbf{D-F}) of the studied system at different volumetric compression regimes. Color indicates the edge length of the simulation (in nm), assuming a cubic simulation box, see Table~\ref{tab:parameters} and Fig.~\ref{fig:postEquil}. Panels \textbf{A}, \textbf{D} correspond to the results obtained with the MARTINI-std force field, while panels \textbf{B}, \textbf{E} and \textbf{C}, \textbf{F} are obtained using the MARTINI-1.5 and MARTINI-2, respectively.
\label{fig:volumetric_time_evolution}
}
\end{figure}

Figure~\ref{fig:volumetric_time_evolution}A-C shows the time evolution of the internal (potential) energy of the entire simulation box with polystyrene computed for the three different force fields employed, see Table~\ref{tab:protocol}. Figures~\ref{fig:volumetric_time_evolution}A-C were obtained for the different sizes of the simulation box, as indicated with color and feature evolution of the internal energy of the system during the interval of 800~ns, which has been chosen to (i) allow polystyrene to adjust to the new box size, i.e. re-equilibrate, and (ii) sample the energy in order to acquire the $E(V)$ dependence (production simulation). The production simulations were assumed to last 375~ns in all cases.

\begin{figure}[!t]
\centering
\includegraphics[width=15cm]{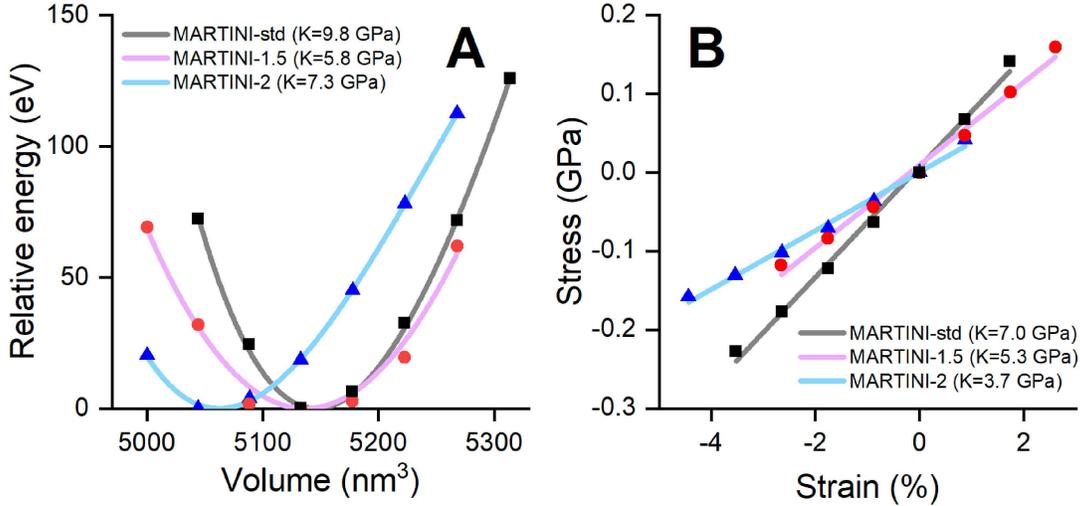}
\caption{
\textbf{First-order elastic properties of the polystyrene sample for volumetric compression.} \textbf{A:} Relative potential energy of the polystyrene sample computed upon volumetric compression computed for the MARTINI-std (black squares), MARTINI-1.5 (red dots) and MARTINI-2 (blue triangles) force fields. The energies were computed as the average values of the energies shown in Fig.~\ref{fig:volumetric_time_evolution}, where only the data that corresponds to the last 375~ns was used for averaging. The values were then computed with respect to the lowest average energy. Lines correspond to the theoretical model, suggested by Birch and Murnaghan, Eq.~(\ref{eq:BirchMurnaghamEnergy}) \cite{Birch1947,Murnaghan1944}. The values in the inset indicate the bulk moduli obtained from the fit for the three force fields considered. \textbf{B:} Stress-strain curve for the volumetric compression of the studied polystyrene sample, calculated from the internal pressures shown in Fig.~\ref{fig:volumetric_time_evolution}. The averaging is performed similarly to \textbf{A}. Data is shown for the three considered force fields, and the resulting bulk moduli are shown in the inset.
\label{fig:volumetric_energy_stressstrain}
}
\end{figure}

The average energies from Figs.~\ref{fig:volumetric_time_evolution}A-C, sampled over the production simulations, allow plotting the relative potential energy of bulk polystyrene computed upon volumetric compression, see Fig.~\ref{fig:volumetric_energy_stressstrain}A. The results of the simulations are shown with symbols, while solid lines correspond to the theoretical model of Birch and Murnaghan, Eq.~(\ref{eq:BirchMurnaghamEnergy}). Fitting the simulation points with Eq.~(\ref{eq:BirchMurnaghamEnergy}) allows establishing the values of the bulk moduli for the three force fields considered, as shown in the inset. The figure shows that the simulated data matches the theoretical curve closely and reveals that the bulk modulus of polystyrene decreases if the van der Waals interaction between the coarse-grained beads is decreased, i.e., the value of $K$ appears lower for the MARTINI-2 force field, as compared to the value obtained in   MARTINI-std. This happens because the stiffness of the polystyrene chains in the considered model is controlled through the parameter $\varepsilon$ of the MARTINI force field equation, Eq.~(\ref{eq:MARTINI}). This parameter is steadily decreased, see Table~\ref{tab:parameters}, for the MARTINI-std$\rightarrow$MARTINI-1.5$\rightarrow$MARTINI-2 force fields shifts, thereby making bulk polystyrene in the case of MARTINI-2 have a lower bulk modulus than in the case of MARTINI-std force field. Note that the value of $K$ obtained for MARTINI-1.5 force field from fitting the average energies with Eq.~(\ref{eq:BirchMurnaghamEnergy}) appears to be lower than in the case of MARTINI-2. This can be attributed to the significant uncertainty of energy, as follows from Fig.~\ref{fig:volumetric_time_evolution}, which makes the obtained volumetric stress values obtained in Fig.~\ref{fig:volumetric_energy_stressstrain}A rather qualitative.

It is important to stress that the density of the bulk polystyrene used in the computational study of volumetric compression appears to be lower than the equilibrium value of 960 kg/m$^3$. This is evident from Table~\ref{tab:protocol} and is also featured in Fig.~\ref{fig:volumetric_energy_stressstrain}A, where the minimum of the bulk polystyrene potential energy appears at a volume that is somewhat lower than 167.2$\times$167.2$\times$167.2 \AA$^{3}$. The approximate density of bulk polystyrene at the minimum of the potential energy is 880~kg/m$^3$, which is still an acceptable value for polystyrene \cite{Zoller1995,rossi2011coarse}. The reason for this lowering of density is because the original long-equilibration was performed at a fixed polystyrene density of 960 kg/m$^3$, which, however, does not match the configuration with the minimum potential energy for the employed force fields. Once the simulation box size changes, it is, therefore, important to perform an additional equilibration to ensure that the system accommodates to the change of its dimensions and redistributes the density uniformly. This has been achieved through the 425~ns long equilibration simulations, see Fig.~\ref{fig:volumetric_time_evolution}.

The obtained bulk moduli could now be compared to experimental values. For polystyrene, one expects to have a volumetric compressibility modulus of about 3~GPa \cite{Mark1999}. This value is, however, notably different from the fitted values shown in Fig.~\ref{fig:volumetric_energy_stressstrain}A. Although the bulk modulus for MARTINI-1.5 and MARTINI-2 is closer to the experimental data, the computed points are expected to have rather significant error bars. As was already pointed out before, the fluctuations of energy in Figs.~\ref{fig:volumetric_time_evolution}A-C make it very difficult to distinguish between the individual energy curves, leading to a conclusion that measuring the energy as the function of deformation gives qualitatively a reasonable description of polystyrene volumetric compression; however, it can not be used for quantitative analysis.

An alternative approach is to compute the hydrostatic pressure of the system, which can be defined as:

\begin{equation}
P=\frac{Nk_{B}T}{V}+\frac{1}{3V}\langle\sum_{i=1}^N\vec{r}_i\vec{F}_i^{int}\rangle.
\label{eq:hydrostatic}
\end{equation}

\noindent
Here $N$ is the total number of particles in the simulation box, $V$ is its volume, $T$ is the temperature of the system, $k_B$ is the Boltzmann factor, $\vec{r}_i$ is the position of a particle with the index $i$, and $\vec{F}_i^{int}$ is the internal force acting on it. The hydrostatic pressure is related to the stress tensor $\hat{\sigma}$ in the following way

\begin{equation}
P=\frac{Nk_{B}T}{V}-\frac{1}{3}\mathop{\mathrm{Tr}}\hat{\sigma},
\label{eq:hydrostatic2}
\end{equation}

\noindent
where $\frac{1}{3}\mathop{\mathrm{Tr}}\hat{\sigma} = \frac{1}{3}(\sigma_{xx} + \sigma_{yy} + \sigma_{zz})$ is the volumetric stress.
The linear elasticity theory suggests \cite{LL7} that the stress tensor $\hat{\sigma}$ is proportional to the strain tensor $\hat{\varepsilon}$, such that

\begin{equation}
\sigma_{ij} = \sum_{kl}C_{ijkl}\varepsilon_{kl},
\end{equation}

\noindent
where $C_{ijkl}$ is the stiffness tensor. For an isotropic medium, it has the form

\begin{equation}
C_{ijkl}=(K + \frac{2}{3}G)  \delta_{ij} \delta_{kl}+G \left(\delta_{il} \delta_{jk}+\delta_{ik} \delta_{jl}\right),
\label{eq:Cijk}
\end{equation}

\noindent
where $K$ and $G$ are the bulk modulus and the shear modulus, respectively. Thus, for any deformation tensor $\hat{\varepsilon}$, the volumetric stress has the form

\begin{equation}
\frac{1}{3}\mathop{\mathrm{Tr}}\hat{\sigma} = K \mathop{\mathrm{Tr}}\hat{\varepsilon},
\label{eq:K}
\end{equation}

\noindent
where $\mathop{\mathrm{Tr}}\hat{\varepsilon} = \delta V/V$ is the volumetric strain.

Figures~\ref{fig:volumetric_time_evolution}D-F shows the time evolution of the hydrostatic (internal) pressure of the studied system at different volumetric compression regimes. The plots illustrate that in all simulations, for the three different force fields employed, the internal pressure saturates quickly with fairly small fluctuations. It is especially remarkable that in the case of the MARTINI-2 force field, it is still easily possible to distinguish between the individual curves of the internal pressure vs. simulation time dependence. Through computing the average values of the internal pressure it is ultimately possible to establish the stress-strain dependency for bulk polystyrene in the case of its volumetric compression. Here one defines:

\begin{eqnarray*}
\nonumber % Remove numbering (before each equation)
\mathrm{volumetric\ stress} &=& P_0 - P\ \ \ \mathrm{(GPa)} \\
\label{eq:stressstrain}
\mathrm{volumetric\ strain} &=& \frac{V - V_0}{V_0}100\%,
\end{eqnarray*}

\noindent
where $V_0$ and $P_0$ are the reference values of the simulation box volume and the associated internal pressure.

Figure~\ref{fig:volumetric_energy_stressstrain}B shows the dependence of the stress-on-strain for the case of a volumetric compression of the studied polystyrene system, calculated from the internal pressures shown in Fig.~\ref{fig:volumetric_time_evolution}D-F. In the case of small deformations, it is expected that stress will depend linearly on strain \cite{Slaughter2002}, and the tangent of this dependency will yield the bulk modulus. Figure~\ref{fig:volumetric_energy_stressstrain}B illustrates that the obtained values for the bulk moduli for MARTINI-std, MARTINI-1.5, and MARTINI-2 force field are different from those shown in the inset to Fig.~\ref{fig:volumetric_energy_stressstrain}A, and appear to be much closer to those expected from experiment \cite{Mark1999}. In particular, the value of $K=3.7$~GPa gives a good agreement, and suggests the at the MARTINI-2 modification of the standard MARTINI force field may lead to some improvements in the computation of the bulk modulus.

%%%%%%%%%%%%%%%%%%%%%%%%%%%%%%%%%%%%%%%%%%%%%%%%%%%%%%%%%%%%%%%%%%
\subsection{Uniaxial compression}
%%%%%%%%%%%%%%%%%%%%%%%%%%%%%%%%%%%%%%%%%%%%%%%%%%%%%%%%%%%%%%%%%%

The second compression regime to consider is uniaxial compression. Equation~(\ref{eq:K}) holds for any deformation and can also be applied to uniaxial deformation of the polystyrene system, yielding the bulk modulus as well. Such a uniaxial compression can also yield other elastic constants, for example, the shear modulus. In the case of the uniaxial deformation, the volumetric strain is equal to axial strain, namely

\begin{equation}
\frac{V-V_0}{V_0} = \frac{L_x-L_0}{L_0}.
\label{eq:axialstrain}
\end{equation}

\noindent
Here $L_x$ and $L_0$ are the deformed and the reference values of the polystyrene box along the compression x-axis.

\begin{figure}[!t]
\centering
\includegraphics[width=15cm]{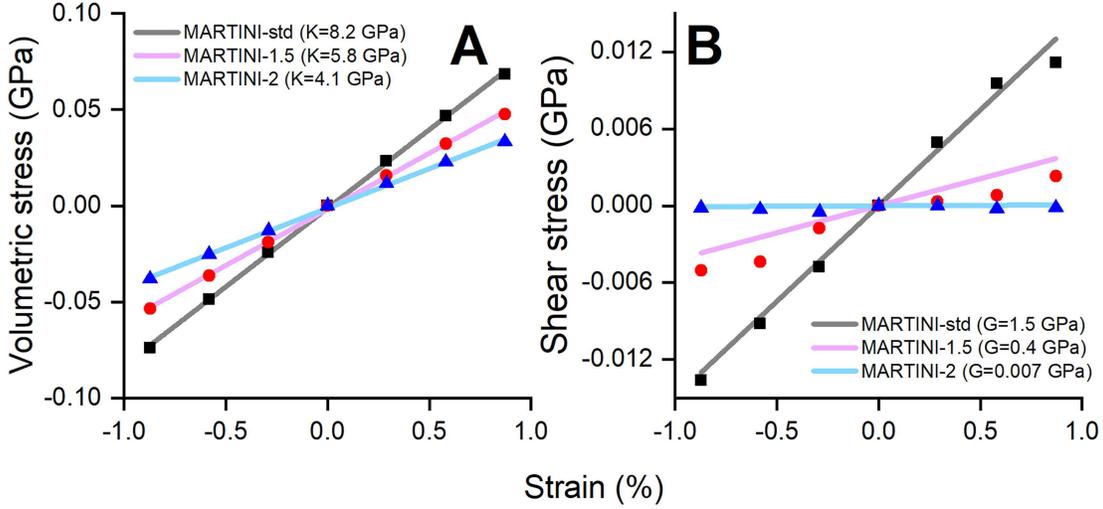}
\caption{
\textbf{First-order elastic properties of the polystyrene sample for uniaxial compression.} Volumetric stress-strain dependencies obtained for the studied polystyrene system upon uniaxial compression (\textbf{A}). Panel \textbf{B} shows the strain dependence of the shear stress for the same system. Results were obtained using the three force field modifications, employed in this study: MARTINI-std (black squares), MARTINI-1.5 (red dots) and MARTINI-2 (blue triangles).
\label{fig:axial_stressstrain}
}
\end{figure}

\begin{figure}[!h]
\centering
\includegraphics[width=15cm]{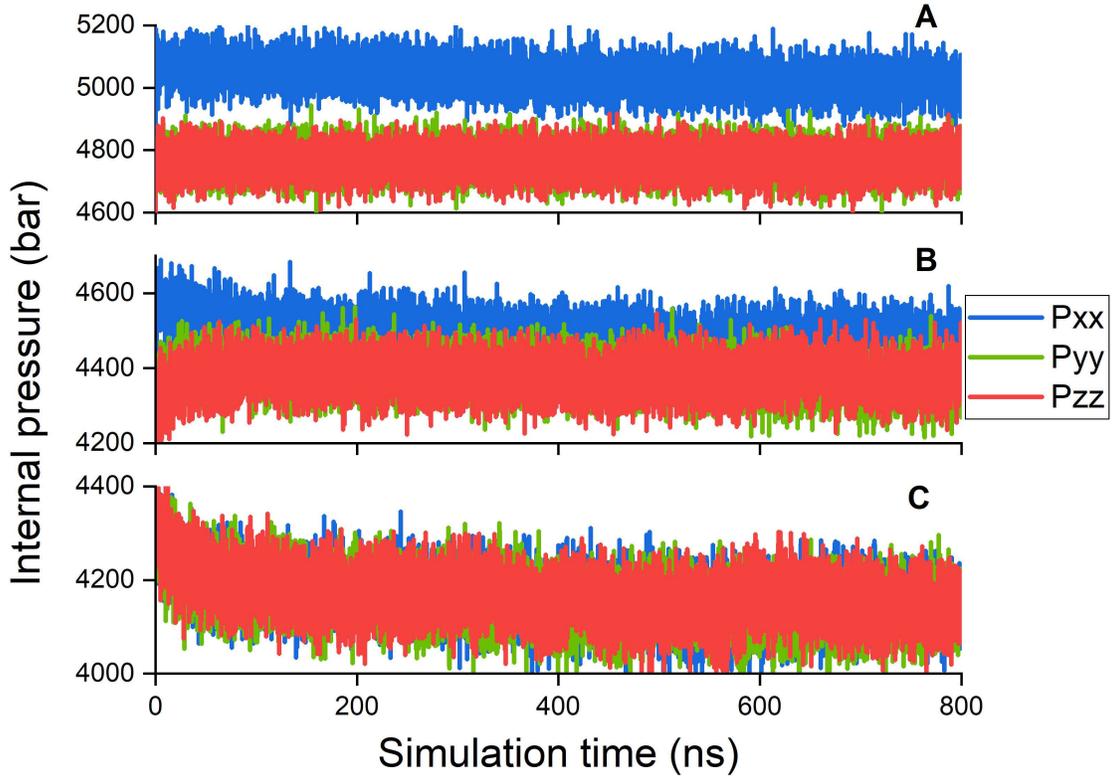}
\caption{
\textbf{Principal components of the pressure tensor.} The time evolution of the principal components $P_{xx}$ (blue), $P_{yy}$ (green) and $P_{zz}$ (red) are shown for the uniaxial deformation of the studied polystyrene system in a box of 17.05~nm$\times$17.2~nm$\times$17.2~nm size, computed for MARTINI-std (\textbf{A}), MARTINI-1.5 (\textbf{B}) and MARTINI-2 (\textbf{C}) force fields.
\label{fig:pressure_assymetry}
}
\end{figure}

The results of volumetric stress-strain dependence obtained in the uniaxial compression simulation are shown in Fig.~\ref{fig:axial_stressstrain}A. In this case, the bulk moduli are 8.2 GPa, 5.8 GPa, and 4.1 GPa for MARTINI-std, MARTINI-1.5, and MARTINI-2 force fields, respectively. These values are 7--17\% higher than those obtained from volumetric compression. This difference can be caused by the local heterogeneity and anisotropy of the simulated polystyrene. It is however remarkable that in the case of a uniaxial compression the value of the bulk modulus steadily decreases for the MARTINI-std$\rightarrow$MARTINI-1.5$\rightarrow$MARTINI-2 force fields, as the fluctuations of the pressure components are smaller, providing more distinct values.

In order to obtain the shear modulus $G$, it is important to consider the pressure as a tensor with different components. The pressure tensor is related to the stress tensor as follows:

\begin{equation}
\sigma_{\alpha\beta} = (P_0)_{\alpha\beta} - P_{\alpha\beta},
\label{eq:presTens}
\end{equation}

\noindent
where $\alpha=(x,y,z)$ and $\beta=(x,y,z)$. In the case of the uniaxial compression along the $x$ direction of the simulation box, there is only one nonzero component of the deformation tensor, $\varepsilon_{xx}$. The stress tensor, in this case, has a diagonal form with the diagonal elements given as

\begin{equation}
\sigma_{xx} = \left(\frac{4}{3}G + K\right)\varepsilon_{xx}, \quad \sigma_{yy} = \sigma_{zz} = \left(-\frac{2}{3}G + K\right)\varepsilon_{xx}.
\label{eq:sigmaXX}
\end{equation}

\noindent
Combining the diagonal components of the stress tensor one obtains:

\begin{equation}
\frac{2\sigma_{xx} - \sigma_{yy} - \sigma_{zz}}{4} = G \varepsilon_{xx}.
\label{eq:shear}
\end{equation}

\noindent
This equation gives an expression for the shear stress, that can be computed from the components in its left-hand side. The results for the shear stress are given in Fig.~\ref{fig:axial_stressstrain}B. The following values for the shear stress are obtained after fitting the simulation data with the linear approximation: 1.5 GPa, 0.42 GPa, 0.007 GPa for MARTINI-std, MARTINI-1.5 and MARTINI-2 force fields, respectively. It is remarkable that the shear modulus for the system simulated with the MARTINI-2 force field becomes close to zero. This happens because the melting temperature of the polystyrene system becomes below 300 K. In this case, the interaction between individual parts of the polymeric chains becomes too weak, and the polystyrene model behaves like a liquid with zero shear modulus.

Additionally, to the bulk modulus and shear modulus, an important elastic parameter of any material is the Poisson ratio. It can be determined using the known values for the bulk modulus $K$ and the shear modulus $G$ as:

\begin{equation}
\nu = \frac{3K - 2G}{6K + 2G}.
\label{eq:Poisson}
\end{equation}

\noindent
The simulation results discussed above allow calculating the Poisson ratio for MARTINI-std, MARTINI-1.5 and MARTINI-2 force fields being equal to 0.4, 0.46, 0.499, respectively.
All the values are a bit higher than the typical value for the polystyrene, being 0.34 \cite{Harper2002,Zoller1995}. It is also remarkable that the Poisson ratio value obtained for MARTINI-2 is close to 0.5, which is the expected value for the liquid state. This result suggests that once the interactions between polymeric chains are significantly reduced (as in the case of the MARTINI-2 force field) the melting temperature decreases.

%%%%%%%%%%%%%%%%%%%%%%%%%%%%%%%%%%%%%%%%%%%%%%%%%%%%%%%%%%%%%%%%%%
\section{Conclusion}
%%%%%%%%%%%%%%%%%%%%%%%%%%%%%%%%%%%%%%%%%%%%%%%%%%%%%%%%%%%%%%%%%%

The present investigation documents an approach on how to obtain elastic properties of polymer-based systems from molecular dynamics simulations. Polystyrene is considered here as a specific case study, which is investigated through the coarse-grained MARTINI force field. The reported results indicate that the conventional MARTINI force field overestimates elastic moduli (both bulk and shear moduli) as compared to experiment. We propose here that the elastic constants could be efficiently tuned through the modifications of the MARTINI force field by reducing the inter-particle van der Waals forces. Two such modifications were studied here, and it was demonstrated that significantly different values of the bulk and shear moduli for the system could be obtained. The simulations have, however, also revealed that the reduction of the van der Waals interaction leads to decreasing of the polystyrene melting temperature: once the van der Waals interactions between the coarse-grained are reduced by a factor of 2 or more, the shear modulus drops to 0, which indicates the melting of the polymeric material for the simulation temperature is below 300 K.

Based on the obtained results we, therefore, conclude that coarse-grained simulations permit performing significantly long molecular dynamics simulations of polymeric systems (here we have considered nearly 1~$\mu$s long simulations), which take slow relaxation processes into account. On the other hand, the physical correctness of the coarse-grained system turns out to be very parameter dependent, and could likely be only considered qualitatively. Indeed, we have demonstrated that both the bulk and shear moduli of polystyrene could be obtained, that would be comparable to the values known from experiment, however, it is not entirely clear how the derived force fields would apply to the modified polystyrene systems, e.g., those doped with nanoparticles. Here a followup investigation is called for, which can root upon the results presented in the present manuscript, but deserves a separate validation and justification.

\section*{Acknowledgements}
Financial support by the Russian Science Foundation(RSF), grant no. 17-72-20201, is greatly acknowledged.
Computational resources for the simulations were provided by the DeiC National HPC Center, University of Southern Denmark.

%\bibliographystyle{unsrt}
%\bibliography{bib/ilia,bib/journals_short,bib/IS_group}

\end{document}